**Ask2Me VarHarmonizer: A Python-Based Tool to Harmonize Variants from Cancer Genetic Testing Reports and Map them to the ClinVar Database**

Yuxi Liu, MS[1,2], Kanhua Yin, MD, MPH[3,4], Basanta Lamichhane, MS[5], John F. Sandbach, MD[6], Gayle Patel, CGC[6], Gia Compagnoni, MD[7], Richard H. Kanak[7], Barry Rosen, MD[7], David P. Ondrula, MD[7], Linda Smith, MD[8], Eric Brown, MD[9], Linsey Gold, DO[9], Pat Whitworth, MD[10], Colleen App, MD[11], David Euhus, MD[12], Alan Semine, MD[13], S. Dwight Lyons, MD[14], Melford Allan C. Lazarte, APRN[14], Giovanni Parmigiani, PhD[2, 15, **], Danielle Braun, PhD[2, 15, **], and Kevin S. Hughes, MD[3, 4, **]

1. Department of Epidemiology, Harvard T.H. Chan School of Public Health, Boston, MA

2. Department of Data Sciences, Dana-Farber Cancer Institute, Boston, MA

3. Division of Surgical Oncology, Massachusetts General Hospital, Boston, MA

4. Harvard Medical School, Boston, MA

5. SHEER Services, Kathmandu, Nepal

6. Texas Oncology, Austin, TX

7. Advanced Surgical Care of Northern Illinois, Advocate Health Care, Barrington, IL

8. New Mexico Comprehensive Breast Care, Albuquerque, NM

9. Comprehensive Breast Care, A Division of Michigan Healthcare Professionals, Troy, MI

10. Nashville Breast Center, Nashville, TN

11. The Breast Health and Wellness Center, Grand Rapids, MI

12. Johns Hopkins Hospital, Johns Hopkins University School of Medicine, Baltimore, MD

13. Newton-Wellesley Hospital, Newton, MA

14. Lyons Care Associates, Kahului, HI

15. Department of Biostatistics, Harvard T.H. Chan School of Public Health, Boston, MA

** co-senior authors

Corresponding author:

Danielle Braun, PhD. Department of Biostatistics, Harvard T.H. Chan School of Public Health, Department of Data Sciences, Dana-Farber Cancer Institute, 677 Huntington Ave, SPH 4th Floor, Boston, MA 02115. Tel: +1-617-632-3654; Fax: +1-617-632-5444; E-mail: dbraun@mail.harvard.edu

**Conflict of Interest**

Disclosures: Dr. Hughes receives Honoraria from Myriad Genetics Veritas Genetics, Advisory Board for Beacon (An RFID Biopsy Marker) and is a founder of and has a financial interest in Hughes Risk Apps, LLC. Dr. Hughes's interests were reviewed and are managed by Massachusetts General Hospital and Partners Health Care in accordance with their conflict of interest policies.

Dr. Parmigiani is a member of the Scientific Advisory Board and has a financial interest in Cancer Risk Apps LLC (CRA). CRA commercializes software for management of patients at high risk of cancer. He is also a member of the Scientific Advisory Board of Konica-Minolta who owns Ambry genetics.

Dr. Braun and Dr. Parmigiani co-lead the BayesMendel laboratory, which licenses software for the computation of risk prediction models. These authors do not derive any personal income from these licenses. All revenues are assigned to the lab for software maintenance and upgrades.

We feel there is no significant overlap with this work.

# ABSTRACT

**PURPOSE:** The popularity of germline genetic panel testing has led to a vast accumulation of variant-level data. Variant names are not always consistent across laboratories and not easily mappable to public variant databases such as ClinVar. A tool that can automate the process of variants harmonization and mapping is needed to help clinicians ensure their variant interpretations are accurate.

**METHODS:** We present a Python-based tool, Ask2Me VarHarmonizer, that incorporates data cleaning, name harmonization, and a four-attempt mapping to ClinVar procedure. We applied this tool to map variants from a pilot dataset collected from 11 clinical practices. Mapping results were evaluated with and without the transcript information.

**RESULTS:** Using Ask2Me VarHarmonizer, 4728 out of 6027 variant entries (78%) were successfully mapped to ClinVar, corresponding to 3699 mappable unique variants. With the addition of 1099 unique unmappable variants, a total of 4798 unique variants were eventually identified. 427 (9%) of these had multiple names, of which 343 (7%) had multiple names within-practice. 99% mapping consistency was observed with and without transcript information.

**CONCLUSION:** Ask2Me VarHarmonizer aggregates and structures variant data, harmonizes names, and maps variants to ClinVar. Performing harmonization removes the ambiguity and redundancy of variants from different sources.

# INTRODUCTION

With the widespread use of high-throughput sequencing and the rapid reduction in cost of genetic testing, the number of identified genetic variants has grown. Clinicians ordering genetic testing receive results on a large number of germline variants for a large number of patients. As new data accumulates, genetic variant interpretations may change: variants of unknown significance (VUS) can change to benign or pathogenic, and even pathogenic variants can change to benign.[1-3] It is therefore of utmost importance for genetic testing labs and physicians to regularly check their existing genetic variant classifications against a publicly available database (e.g., ClinVar) and other sources, to keep their interpretations up to date and inform patients of changes.

Three major challenges must be overcome to accomplish this task. First, it is not feasible for labs or clinicians to manually map variants to existing variant databases, as the number of variants is extremely large. Second, inconsistencies exist in variant representations, especially in genetic reports released before the Human Genome Variation Society (HGVS) nomenclature guidelines[4] became standard. Multiple naming conventions exist and the same variant can be denoted in several ways by different labs. For example, BRCA1 (NM_007294.3) variant c.68_69delAG is also recorded as 185delAG or 187delAG; BRCA1 (NM_007294.3) variant c.181T>G is also recorded as 300T>G. No simple approach exists to map existing naming conventions to a single harmonized name for each variant. Third, recording of variants is usually not in an accessible structured format limiting the usefulness of existing variant mapping tools. Genetic results are recorded in a variety of file formats, most of which, unfortunately, are not structured (e.g. portable document format (PDF) reports or electronic health record (EHR) notes). Even when structured data is recorded, manual entry may result in recording the same variant in slightly different notations (e.g., c.68_69del, c68_69del, 68_69del). This variation can render the variant unrecognizable by ClinVar or other databases.

To address these challenges, we developed a Python-based tool, Ask2Me VarHarmonizer, that can be used to harmonize the different naming conventions of a unique variant into a single standardized name, automatically map it to a public variant archive, ClinVar, and return the naming and classification information. We applied this tool to a pilot variant dataset of more than 7000 variant submissions collected from 11 clinical sites.

**MATERIALS AND METHODS**

**Design of Ask2Me VarHarmonizer.** Our implementation is in two steps: 1) curating and preprocessing the original variant entries; 2) harmonizing variants and mapping to ClinVar. The workflow is summarized in Fig 1 and described next through an example dataset.

*Step 1: Variant curating and preprocessing*

We collected original variant entries from participating practices using a common data collection form, with four required fields: "practice", "gene", "original variant name", and "original lab-reported classification", and four optional fields: "protein change", "transcript ID", "lab", and "test year". Definitions and examples for each field are in Table 1.

The preprocessing steps include: (1) removal of incomplete variant entries: entries with missing or invalid information in at least one of the four required fields were removed (e.g. no data or "0" in the "original variant names" field, or no data or "?" in the "original lab classification" field); (2) optional filtering of genes according to user defined criteria: this allows focusing on subsets of interest (e.g. cancer susceptibility genes).

Variants from lab reports are usually classified into five categories: benign, likely benign, VUS, likely pathogenic, and pathogenic. First, we consolidated the variant classifications into three categories (referred to as "lab-reported classifications"): benign, VUS, and pathogenic. This was done to emphasize major differences in classification for the same variant, (e.g., pathogenic and VUS) and de-emphasize minor differences, which are likely to have less clinical significance (e.g., benign and likely benign). Details regarding the classification consolidation are provided in Table S1. Next, we removed duplicate entries, defined as any two variants having identical data in all the following fields: 1) "practice", 2) "gene", 3) "original variant name", and 4) "lab-reported classification".

***Step 2: Variant harmonizing and mapping to ClinVar***

We next mapped the variants to ClinVar to determine a single name for variants reported in multiple ways. We used the Entrez Direct application programming interface (API) to retrieve ClinVar[5] data.

The entire mapping process includes up to four mapping attempts, described below.

**First attempt:** A combination of "gene", "original variant name", and "transcript ID" (if available) was submitted to the API. If ClinVar uniquely recognizes this variant, a response is obtained. The preferred ClinVar variant name was stored in a new field called "master variant name". Similarly, the corresponding ClinVar classification was also added and stored as a new field titled "ClinVar classification" (Fig 1, M1). **Second attempt:** For variants not mapped in the first attempt, the tool corrected the "original variant name" and stored the resulting correction in a new field labeled "corrected variant name". In this process, the HGVS nomenclature was used as the standard naming convention. Corrections include but are not limited to (1) correcting basic typographical errors and omissions; (2) standardizing the use of "c." (or "p.") before each variant name; and (3) harmonizing deletions and insertions. For example, use of "c." before DNA sequence varies, with "original variant name" sometimes listed as "c." or omitting "c.". For consistency, we confirmed or added "c." before each variant name (e.g., correct "original variant name" for BRCA2 variant from 6895A>G to "corrected variant name": c.6895A>G). We also harmonized deletions to the HGVS convention (e.g., changed "original variant name" for MSH6 variant from c.3699del4 to "corrected variant name": c.3699_3702del). We then submitted a combination of "gene", "corrected variant name", and "transcript ID" (if available) to the API. If ClinVar uniquely recognized this variant, a response was obtained and recorded as described in the first attempt (Fig 1, M2). **Third attempt:** For variants not mapped in the first or second attempts that have information on "protein change", we submitted the combination of "gene", "corrected variant name", "protein change" and "transcript ID" (if available) to the API. If ClinVar uniquely recognized this variant, a response was obtained and recorded as described in the first attempt. Variant entries without "protein change" information were considered unmappable, and their "corrected variant names" were stored in the "master variant names" field (Fig 1, M3). **Fourth attempt:** For those variants that had "protein change" information but were unmappable in the third attempt, we removed the "corrected variant name" from the query and attempted to remap them with the combination "gene", "protein change", and "transcript ID" (if available). Some "corrected variant names" may have been inaccurate and removing this field from the submission allowed us to attempt to map these variants. "master variant names" and "ClinVar classifications" were stored for the mappable variants, while the unmappable ones were considered unmappable, and their "corrected variant names" were stored in the "master variant names" field (Fig 1, M4).

Eventually, we populate the "master variant names" field for all variants, either from ClinVar or from "corrected variant names". Code for this mapping procedure is available upon request.

**Evaluating the impact of transcript reference sequence.** As the transcript reference sequence associated with a variant may not always be available, we validated the impact of missing transcript reference sequences, by selecting a subset of variants with known "transcript ID", and comparing the mapping results with and without "transcript ID".

**Identifying unique variants and assessing multiple names.** We defined a unique variant as a variant entry with unique "gene" and "master variant name" combination. Some unique variants appeared multiple times in the dataset, with differing "original variant name" (e.g. "CHEK2: c.470T>C" had two "original variant names": "c.470T.C" in one practice and "c.470T)C" in another). We refer to these variants as those having multiple names. Unique variants with multiple names within a single practice are referred to as having within-practice multiple names (e.g. "MUTYH: c.1187G>A" has "original variant names": "1187G>A", "c.1187G)A", and "c,1187G>A" in the same practice).

**Evaluating the association between mappable rate and total number of variants in ClinVar.** To evaluate the association between the mappable rates for unique variants (proportion of unique mappable variants of a gene) and the total number of variants in ClinVar, we graphed the mappable rate versus number of variants in ClinVar for all genes and performed a Poisson regression.

**Application.** We collected a pilot dataset of variants from various clinical practices. Eleven clinical practices participated in this pilot, including Advanced Surgical Care of Northern Illinois; Advocate Health Care, Barrington, IL; Bermuda Cancer Genetics and Risk Assessment Clinic, Bermuda; The Breast Health and Wellness Center, Grand Rapids, MI; Comprehensive Breast Care, A Division of Michigan Healthcare Professionals, Troy, MI; Johns Hopkins Hospital (Euhus practice), Baltimore, MD; Massachusetts General Hospital (Hughes practice), Boston, MA; Nashville Breast Center, Nashville, TN; New Mexico Comprehensive Breast Care, Albuquerque, NM; Newton-Wellesley Hospital, Newton, MA; Texas Oncology, Austin, TX.

We applied the developed Ask2Me VarHarmonizer to this pilot dataset, limiting the analysis to the 49 most commonly tested germline cancer susceptibility genes and/or genes with more than 5 variants reported in the pilot dataset: APC, ATM, BAP1, BARD1, BLM, BMPR1A, BRCA1, BRCA2, BRIP1, CDH1, CDK4, CDKN2A, CHEK2, DICER1, DIS3L2, EPCAM, FANCC, FH, FLCN, HOXB13, MEN1, MLH1, MRE11, MSH2, MSH6, MUTYH, NBN, NF1, NF2, PALB2, PMS2, POLD1, POLE, PTEN, RAD50, RAD51C,

RAD51D, RB1, RECQL4, RET, SDHA, SDHB, SMAD4, SMARCA4, STK11, TP53, TSC1, TSC2, and VHL.

The Dana-Farber Cancer Institute (DFCI) Institutional Review Board (IRB) has determined this study does not constitute human subjects research, and consequently formal IRB review is not warranted (DFCI protocol ID: 19-101).

**RESULTS**

**Variant preprocessing.** In all, we collected 7496 variant entries in 132 genes from 11 practices (test period: 1996--2019). We removed variant entries lacking information or contained invalid information in any required field (n=47). We also excluded an additional 530 variant entries in genes other than the 49 of interest, and 892 variant entries with data in the combination "practice", "gene", "original variant name", and "lab-reported classification". A total of 6027 variant entries (80.4% of the total submission) entered the mapping process (Fig 2, Step 1).

**Variant harmonizing and mapping results.** The 6027 variant entries selected in step 1 underwent four attempts at variant harmonization and mapping (Fig 2, Step 2). Among the 437 variants that mapped to ClinVar in the second attempt, 423 were corrected by standardizing the use of the initial "c." and "p." and 77 were corrected by standardizing the format of deletions or insertions. The date of access to ClinVar was April 17, 2019.

Overall, 4728 (78.4%) of the 6027 variant entries were eventually mappable to ClinVar, with their corresponding ClinVar names stored in the "master variant names" field. For the remaining 1299 (21.6%) unmappable variant entries, their "corrected variant names" were stored in the "master variant names" field.

**Impact of the transcript reference sequence.** Among the 6027 variant entries, transcript reference sequence information was available for 3206 (53.2%) variant entries. By comparing the mapping results for these 3206 variants with or without "transcript IDs", we found that for 98.6% (3162 out of 3206) of the variants we obtained the same mapping results, suggesting our mapping results were robust even in the absence of transcript reference sequence.

**Unique variants and multiple names.** Among the 6027 variant entries, there were 4798 variants with unique "gene" and "master variant name" combinations (i.e., unique variants): 3699 (77.0%) mappable, and 1099 (23.0%) unmappable. (Fig 3). The five genes with the largest number of unique variants were: BRCA2 (n = 601), ATM (n = 440), BRCA1 (n = 421), APC (n = 265), and MSH6 (n = 237). The number of unique variants for each gene is summarized in Table S2. Among all unique variants, 474 (9.9%) were found at multiple practices (373 at 2 practices, 73 at 3 practices, and 28 at more than 3 practices). 427 (8.9%) unique variants had multiple names, 343 of which had within-practice multiple names.

**Association between mapping rate and ClinVar variants count for each gene.** With the exception of EPCAM, all genes had a mappable rate greater than 0.6 (Fig 3). The mappable rate and total number of variants in ClinVar for each gene are shown in Fig 4 (we excluded EPCAM as 0 out of 6 EPCAM variants were mappable).

We fit a Poisson model regressing the total number of variants in ClinVar against the mapping rate for the remaining 48 genes. We found a significant negative association. The coefficient for the mapping rate is -2.68 (95% CI: -2.74, -2.61), implying that a 0.1 increase in mapping rate, corresponds on average to a decrease by 0.268 in the log count of the total number of variants in ClinVar.

**DISCUSSION**

We developed a python-based tool, Ask2Me VarHarmonizer, that harmonizes the naming of variants collected from clinical practices and maps these variants to ClinVar. We applied this tool to over 7000 variant reports collected from 11 clinical practices. Of the 6027 variants in 49 genes of interest, 4728 (78.4%) were successfully mapped to ClinVar. A total of 4798 unique variants were eventually identified, and 8.9% of them (n = 427) were found to have multiple names.

Although there are many well-developed tools (Mutalyzer[6], hgvs Python package[7], VariantValidator[8], etc.) that manipulate and validate naming of variants according to HGVS guidelines[4], all of them require complete and correct information on variants, including transcript reference sequence and sequence change in HGVS format for parsing (e.g. "NM_000038.5:c.3920T>A"). However, in practice, variant data recorded by clinicians is not always well formatted and the transcript reference sequences are not always easily found in genetic testing reports, especially for variants tested before the standardized format was widely used. Hence, there is no simple, automatic way for clinicians to submit lists of variants that have been collected over a period of years to public databases and successfully acquire an updated interpretation of clinical significance for each variant.

Our tool, which was developed specifically for handling variants collected from germline genetic testing reports, can address problems of multiple naming conventions and recording errors. There are two sources of variability in naming for variants that clinicians have collected from lab reports. First, different labs may report different names for a unique variant, and the same lab may report multiple names for a unique variant over time. This may be the result of using different reference transcripts (e.g. NM_007294.3 (BRCA1): c.68_69delAG is also known as NM_007300.3 (BRCA1): c.66_67delAG) or different formatting standards. Second, clinicians may record a unique variant from two identical reports differently. This is very common in the variant data we collected (e.g. NM_007294.3 (BRCA1) variant c.68_69delAG was sometimes denoted as c,68_69delAG). These are considered typographical errors, and, other than this tool, none of the existing above-mentioned tools is equipped to correct such errors. During our variant name standardization step, before the second mapping attempt, our tool explicitly corrects various typographical errors that are commonly seen in clinicians' records (e.g. correct CHEK2 (NM_007194.3) variant c.470T.C to c.470T>C) and harmonizes the variability from lab reports (e.g. correcting APC (NM_000038.5) variant 426delAT to c.426_427delAT).

Our results from the pilot dataset demonstrated that 69.3% (n = 4178) of variants can be mapped to ClinVar in the first attempt of our four-attempt mapping tool (Fig 2). For the remaining unmappable variants, their "original variant names" are likely to be nonstandard and therefore not recognized by ClinVar directly. Our tool automatically standardized the variant names to resolve the variabilities in variant naming, then mapped those standardized names ("corrected variant name") to ClinVar. We found that 7.3% (n = 437) of the total variants in the pilot dataset were mappable in our second attempt, showing that those variants were originally recorded in a less accurate, but correctable, format. For the remaining unmappable variants, we leveraged their protein-level change information, if available. Using this information, in addition to the DNA-level change, can further improve precision and accuracy by mapping 1.4% (n = 86) additional variants. Since the remaining unmappable variants may have severe typographical errors, incorrect formats, or lack information in the variant name, but still have well-formatted protein-level change information, by removing the "corrected variant name" from the query, an additional 0.4% (n = 27) variants became mappable.

For the mappable variants, we were able to further standardize the naming and to aggregate unique variants to single ClinVar preferred names; for the unmappable variants, especially those reported multiple times by the same or different practices, our database can serve as a proxy for ClinVar, and eventually we hope to encourage the enrolled clinical practices to submit these variants to ClinVar. In our pilot dataset, 1299 variant entries could not be mapped to ClinVar. Possible reasons for unsuccessful mapping include: 1) The variants were not submitted to ClinVar; 2) Ambiguity in the variant name that could not be uniquely recognized by ClinVar. e.g. "PMS2 (NM_000535.6): c.2T>?"; 3) The variant name was not recorded in standard HGVS format: For example, BRCA1 (NM_007294.3): del exons 21-24; 4) There were typographical errors that cannot be corrected, e.g. "," in the nucleotide location.

In our study, we found that only half of the variant entries had transcript reference sequence information, though our validation test demonstrated that 98.6% of the time we obtained the same mapping results with or without the reference sequence. Our analysis showed that 9% of the unique variants we identified had multiple names. These differences in recording of variants make data consolidation challenging. Multiple patients in the same practice were found to have the same variant, but named differently. It would be ideal for practices to know which patients have the same variant in a gene. Linking these individuals (if related) can help find relatives that might otherwise be missed and help determine which bloodline the variant came through. For future research, obtaining the maximum number of people with a variant from each practice will aid the determination of classification and penetrance.

Our study also showed an inverse association between the mappable rate and the total number of variants in ClinVar. We conjecture that this inverse association may be associated with incomplete variant coverage in ClinVar. Although several commonly tested genes have been sequenced for decades, many variants, especially from legacy variants reports, were not submitted to ClinVar.

Our tool will be especially useful for low-resource clinical practices seeking to re-evaluate their variant reports at no cost. We will incorporate it into our clinical decision support tool, Ask2Me[9] for clinician practices to submit lists of variants from genetic testing, and obtain a report with harmonized variant names, updated classifications from ClinVar, and comparison results with other practices. These comparison results will be important in understanding the clinical significance of the variants and may help variant reclassification. The discordance between classifications across practices is also important and has been discussed elsewhere (Yin K et al. Legacy Genetic Testing Results for Cancer Susceptibility: How Common Are Conflicting Classifications in a Large Variant Dataset from Multiple Practices?, unpublished).

The intent of our method is not to provide final adjudication of variant names or classifications but rather to identify probable variant names and discrepancies in classification and report back to each practice, with the expectation that each practice will then work with their lab or labs to adjudicate classifications. We hope that this will provide a feedback mechanism to increase our accuracy over time.

Remaining challenges in our computational pipeline are mapping capacity and accuracy given the available variant data. We are constantly improving our mapping algorithm and hope to encourage the participating practices to add unsubmitted variants to ClinVar.

The Ask2Me VarHarmonizer is designed to harmonize naming of variants across clinical practices and to map them to ClinVar. We expect that this tool will allow clinicians to harmonize their variant reports in an effective and efficient manner, to access the most up-to-date information from ClinVar, to identify discordant results that individual practices can adjudicate with their labs, and to record the variants in a more consistent and structured format, ultimately providing better care for patients. The current system that uses a variety of naming conventions runs the risk of inappropriately managing patients. Harmonization of naming conventions with our tool will increase the likelihood that a given patient with a given variant will be appropriately managed.

## ACKNOWLEDGMENTS

We thank Ann Adams (Department of Surgery, Massachusetts General Hospital) for her editorial assistance and Chen Huang for his assistance in editing the figures.

**FIGURE LEGENDS**

**Fig 1.** Workflow and example input dataset of Ask2Me VarHarmonizer. M1-M4 correspond to the four mapping attempts. P1 is the preprocessing step; P2-P4 are the three processes before M2-M4, respectively. The example input dataset only shows part of the input fields.

**Fig 2.** Flow chart of variant preprocessing and mapping results.

**Fig 3.** Summary of mapping results by gene. The x-axis includes the 49 genes in our dataset. The top panel represents the percentage of unique mappable variants for each gene; the bottom panel represents the count of unique variants for each gene.

**Fig 4.** Relationship between the proportion of unique mappable variants in our database and count of ClinVar total variants for each gene (excluding EPCAM). Each dot represents a gene, there are 48 genes plotted.

**Table 1.** Data collection table

| Field | Description | Example |
| --- | --- | --- |
| **Practice** | The institution, hospital, medical system or practice submitting variants | Bermuda Cancer Genetics and Risk Assessment Clinic |
| **Gene** | HGNC gene symbol[10] | APC |
| **Original variant name** | Variant name recorded by that practice, mostly as HGVS DNA-level change | c.665A>T |
| **Original lab-reported classification** | Classification reported by lab | VUS |
| **Protein change**[a] | Effect of the variant, mostly as HGVS protein-level change | p.Gln222Leu |
| **Transcript ID**[a] | Transcript reference sequence: RefSeq[11] or Ensembl Transcript[12] | NM_000038.5 |
| **Lab**[a] | Lab that performed the test | Invitae |
| **Test year**[a] | Year of test | 2017 |

[a] Optional fields

**Fig 1.**

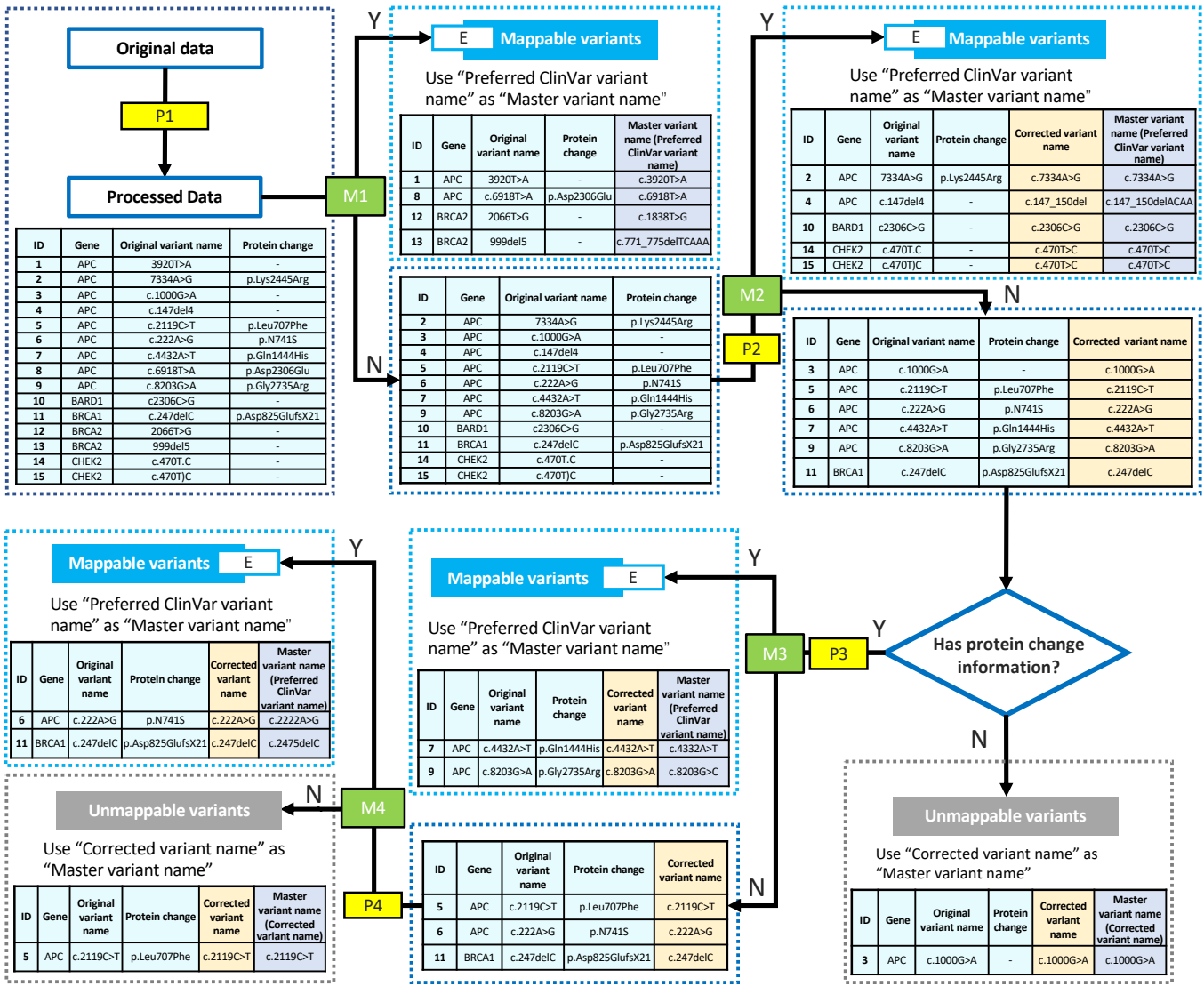

**Fig 2.**

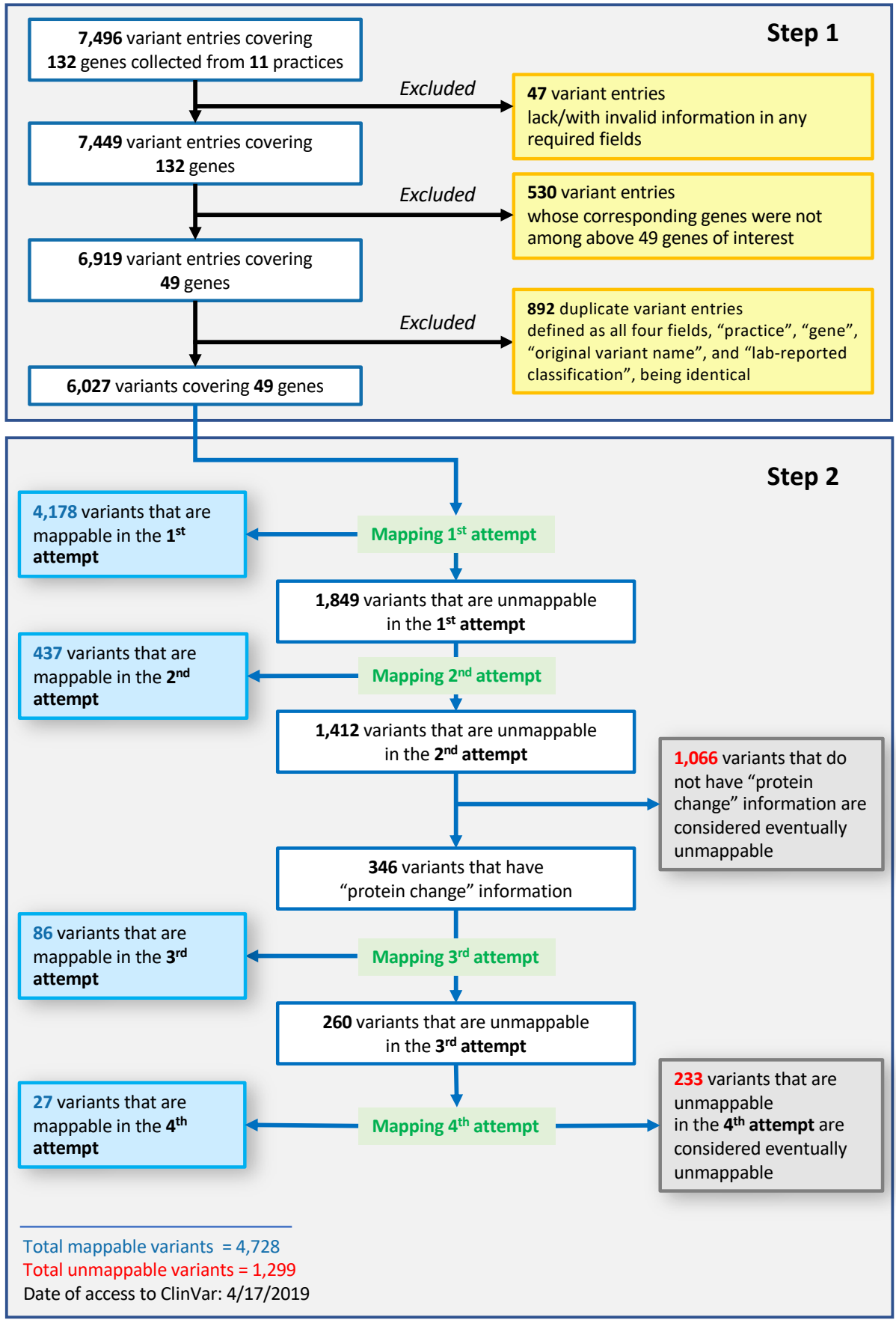

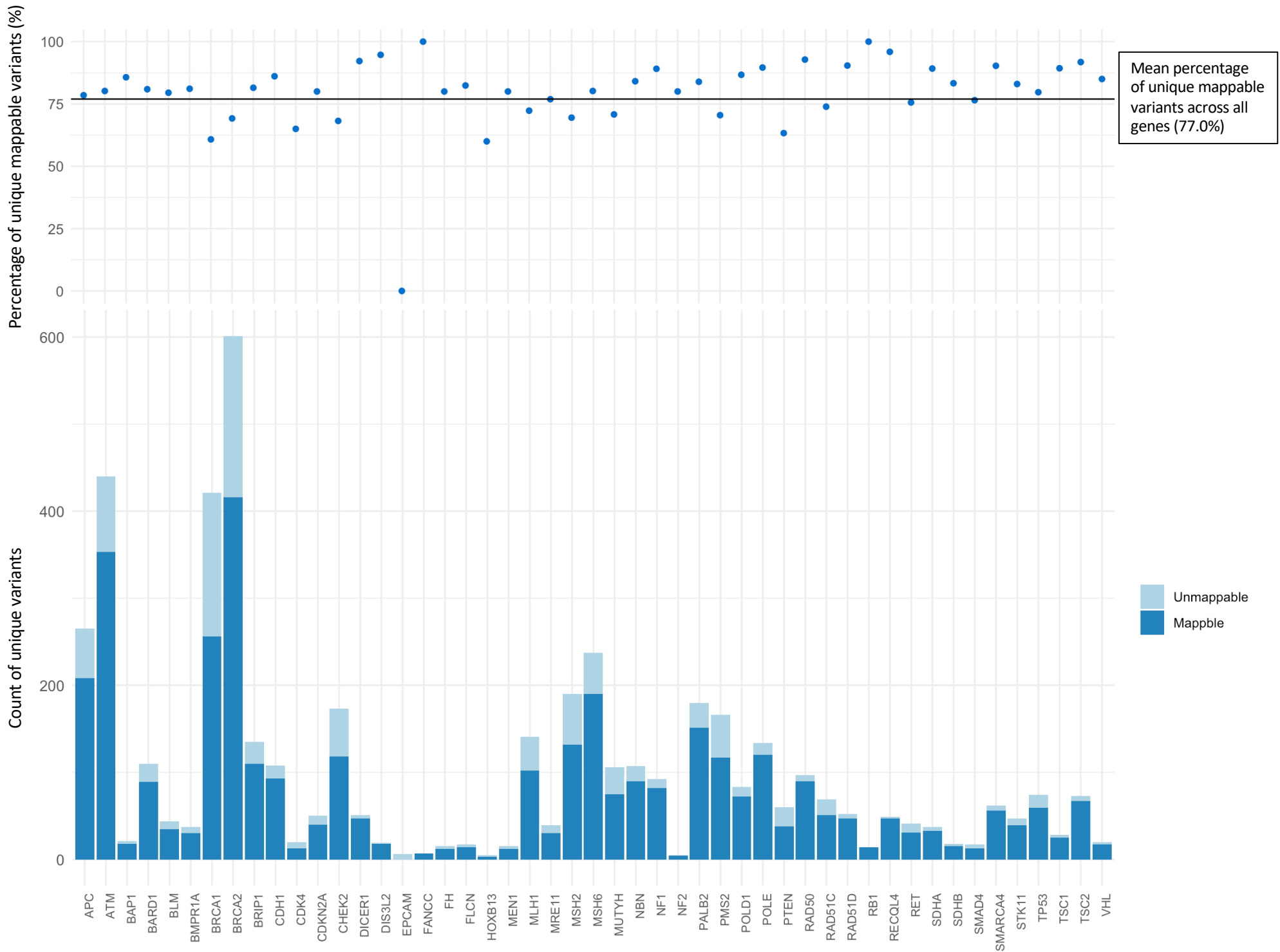

Fig 3.

**Fig 4.**

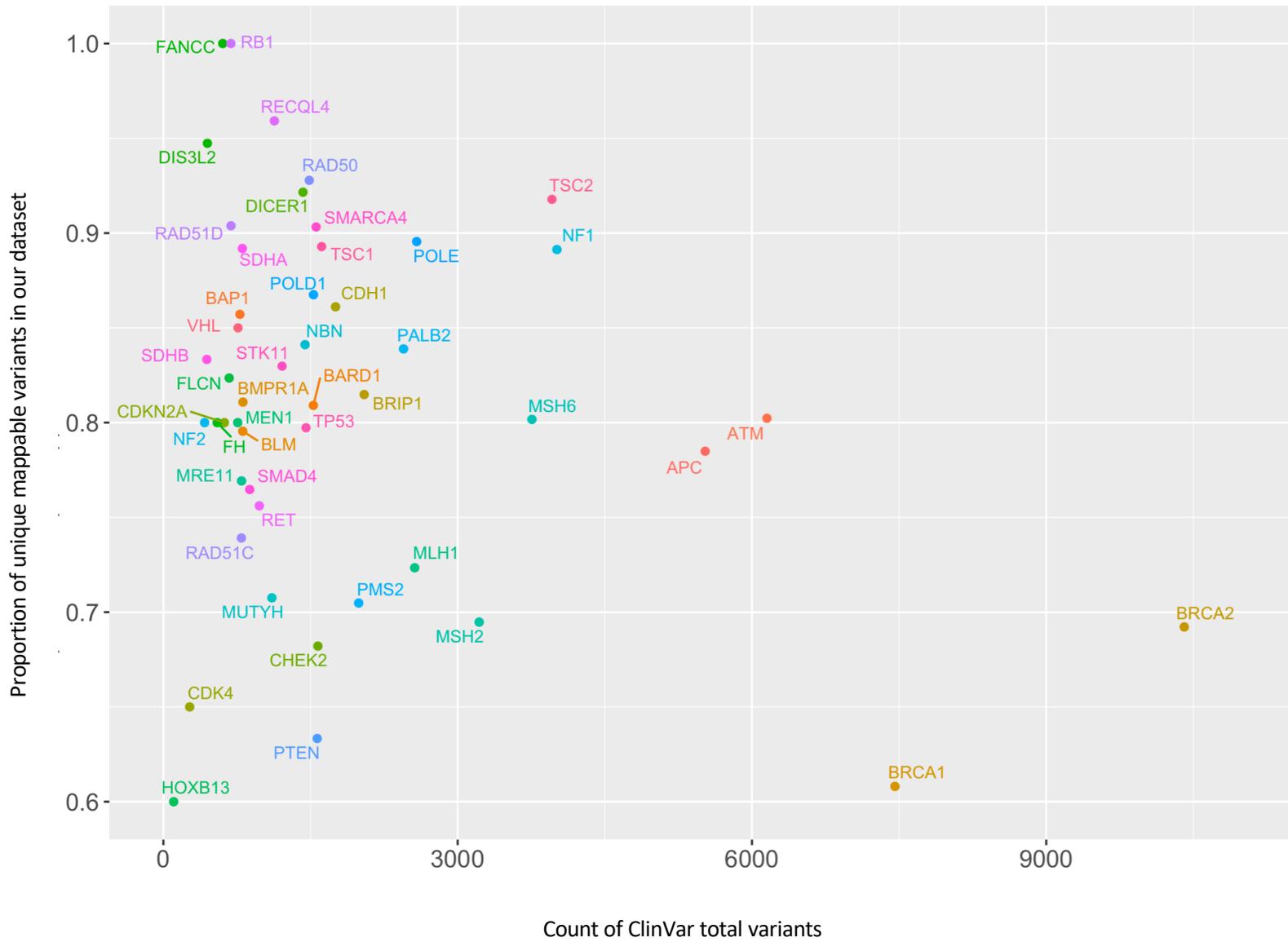